# Fundamentals of Vacuum Physics and Technology

*E. Al-Dmour*
MAX IV laboratory, Lund, Sweden

**Abstract**
A good vacuum performance in particle accelerators is vital for their successful operation. The reasons why a good vacuum is needed, together with the basics of vacuum physics and technology and the main gas sources inside the vacuum system, are presented here.

## 1 Introduction

### 1.1 Vacuum definition

The term 'vacuum' can be defined as the state in which a space has less matter than its surrounding. According to ISO 3529/1, a vacuum is defined as the state at which the pressure or the gas density is less than that of the surrounding atmosphere. The lowest pressure on the surface of the Earth is 300 mbar, which occurs on Mount Everest; accordingly, any pressure below 300 mbar is considered vacuum.

According to the previous definition, a vacuum is actually a pressure (p) which is defined as the force (F) exerted perpendicularly to the surface (A):

$$p = \frac{F}{A}. \tag{1}$$

The gas molecules inside a container will exert a force on the container wall when an impact occurs, so a higher pressure occurs for the same force if the surface area is smaller.

The SI unit for pressure is the Pascal (Pa), which is equivalent to N.m$^{-2}$. One Pa is defined as the pressure exerted by one Newton of force perpendicularly over 1 m$^2$ of surface area. Other units which are commonly used are the mbar and Torr, Table 1 shows the conversion factors of the main units used in vacuum technology.

Table 1: Conversion table of the main pressure units

|  | Pa | mbar | Torr | atm |
|---|---|---|---|---|
| **Pa** | 1.00 | 0.01 | $7.50 \times 10^{-3}$ | $9.87 \times 10^{-6}$ |
| **mbar** | 100.00 | 1.00 | 0.75 | $9.87 \times 10^{-4}$ |
| **Torr** | 133.32 | 1.33 | 1.00 | $1.32 \times 10^{-3}$ |
| **atm** | $1.01 \times 10^5$ | $1.01 \times 10^3$ | 760.00 | 1.00 |

### 1.2 Vacuum classification and ranges

Vacuum can be classified into several ranges depending on the pressure value. The vacuum range varies from the low vacuum range at 33 mbar of pressure, to the extreme high vacuum (XHV) when the pressure is below $1 \times 10^{-12}$ mbar. The vacuum ranges, together with typical applications as classified by the American Vacuum Society in 1980, are shown in Table 2 [1]. The very high vacuum (VHV) and ultra-high vacuum (UHV) ranges are the usual ranges at which particle accelerators operate. For example, Fig. 1 shows the accelerator complex of CERN, together with the operational/design pressure of its accelerators. As seen from the figure, there are various operational pressures, the majority of which are in the UHV range [2].

**Table 2:** Vacuum ranges and typical applications

| Vacuum Range | Pressure Range (mbar) | Typical applications |
| --- | --- | --- |
| Low | $33 < p < 1.0 \times 10^3$ | Vacuum cleaner, mechanical handling, vacuum forming |
| Medium | $1.0 \times 10^{-3} < p < 33$ | Vacuum drying, vacuum freeze (food industries) |
| High (HV) | $1.0 \times 10^{-6} < p < 1.0 \times 10^{-3}$ | Production of microwave, light bulbs, vapour deposition. |
| Very high (VHV) | $1.0 \times 10^{-9} < p < 1.0 \times 10^{-6}$ | Electron microscopes, X-ray and gas discharge tubes, electron beam welding |
| Ultra-high (UHV) | $1.0 \times 10^{-12} < p < 1.0 \times 10^{-9}$ | Particle accelerators, space simulators, material research, semiconductors |
| Extreme high (XHV) | $P \leq 1.0 \times 10^{-12}$ | Particle accelerators, space simulators, advanced semiconductor devices |

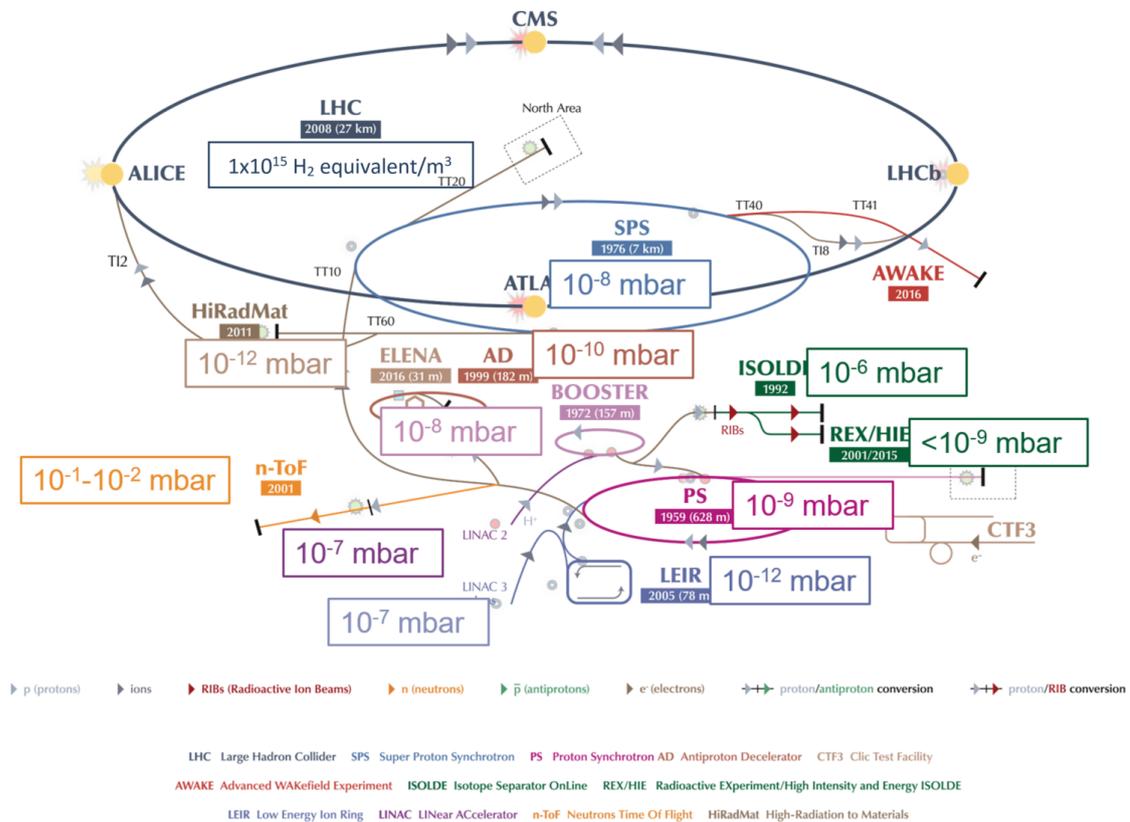

**Fig. 1**: The CERN accelerator complex and the operational pressure of each accelerator

## 2 Vacuum in particle accelerators

The main reason why a good vacuum is essential in particle accelerators is to reduce the interaction between the gas molecules and the circulating beam as those interactions could cause reduction in the beam quality. In summary this could lead to [3]:

(i) Reduction in the lifetime of the stored beam due to elastic and inelastic scattering, this could require frequent injection to compensate for the reduction in the beam current, affecting the experiments being performed (for example increasing the background of the measurements or affecting the size of the beam; or forcing the experiment to stop while injection is being performed due to radiation safety).
(ii) Reduction of the luminosity of the beam.
(iii) Reduction of the bunch intensity.
(iv) Instabilities: such as transverse instabilities driven by ionized molecules, and this in turn could cause beam blow up.
(v) Increase in the induced radioactivity which could have effects on the material such as induced corrosion or radiation damage to instrumentation and also could have an effect on personnel safety due to excess bremsstrahlung radiation.
(vi) Excess noise in the detectors: the interaction between the beam and the gas molecules induces a background to the experiment due to direct effects (non-captured particle interaction with the detectors) or indirect effect due to nuclear cascade.
(vii) Risk of quench in superconductive magnets.

The vacuum is also needed for other reasons:

(i) To avoid contamination of the beamline optics.
(ii) To avoid electrical discharge in high voltage devices such as radio frequency (RF) cavities [4].
(iii) To decrease the heat loss and to provide thermal isolation in cryogenic devices.
(iv) The vacuum chamber can act as an envelope to the circulating beam and will reduce the impedance of the machine.
(v) The vacuum chamber could be used to remove the power from the synchrotron radiation or image current or other sources of power generated by the beam.

## 3 The basics of vacuum physics

A gas in a container can be characterized by three state variables, the volume (V), the pressure (p), and the temperature (T).

The amount of gas can be expressed in different ways: the mass (m), particle number (N), or amount of substance (ν) [5].

The amount of gas is best described by the amount of substance or density rather than by the mass [kg] or particle number (N); the mass of a gas in the field of a vacuum is very small and difficult to quantify.

A more practical way to describe the amount of gas is by using the amount of substance ν [mol], which is a scaling of the amount of a substance to a reference quantity, which is in this case $N_A$:

$$\nu = \frac{N}{N_A}. \tag{2}$$

$N_A$ refers to Avogadro's constant, which is defined as the number of gas particles in one mole; the number of particles in one mole equals $6.02 \times 10^{23}$ particles.

There are other ways to describe the amount of gas:

(i) The mass density (ρ) [kg.m$^{-3}$]: which is defined as the mass divided by the volume which the gas occupies homogenously:

$$\rho = \frac{m}{V}. \tag{3}$$

(ii) Number density (n) [m$^{-3}$]: is the number of particles over the volume:

$$n = \frac{N}{V}. \tag{4}$$

(iii) Molar mass (M) [kg.mol$^{-1}$]: which is the ratio of the mass to the amount of substance:

$$M = \frac{m}{v}. \tag{5}$$

## 3.1 Ideal gas laws

The ideal gas law is the equation of state of the gas, and it is based on some assumptions: the molecules are in a constant state of motion and that motion is in all directions, the volume where the gas is located has a very large number of particles, the gas molecules are assumed to be very small spheres where the distance between them is very large compared to their own diameter, and the molecules exerts no forces on each other unless they collide and that collision is elastic [6].

### *3.1.1 Boyle-Mariotte law*

This describes the relationship between the volume and the pressure of a fixed amount of gas at constant temperature. Boyle and Mariotte found that the product of the pressure and volume is always constant for a given mass at constant temperature.

$$p.V = constant. \tag{6}$$

### *3.1.2 Charles's law*

This describes the relationship between the volume and temperature of a fixed amount of gas at constant pressure. Charles determined that the volume of an ideal gas is directly proportional to its temperature for a given mass at constant pressure.

$$\frac{V}{T} = constant. \tag{7}$$

### *3.1.3 Gay-Lussac's law*

This describes the relationship between the pressure and temperature of a fixed amount of gas for a fixed volume. Gay-Lussac found that the pressure of an ideal gas is directly proportional to its temperature for a given mass at constant volume.

$$\frac{p}{T} = constant. \tag{8}$$

### *3.1.4 Avogadro's law*

This describes the relationship between the volume and the amount of gas at fixed pressure and temperature. Avogadro's law states that the volume of an ideal gas is directly proportional to the number of molecules it contains, given that the pressure and temperature are constant.

$$\frac{V}{n} = constant. \tag{9}$$

In other words, two containers with the same volume at the same pressure and temperature should have the same number of particles.

*3.1.5  Combined or ideal gas law*

The combined gas law gives the relation between the volume, temperature, pressure and amount of gas and it is obtained by combining the laws above [5].

$$\frac{p.V}{T} = constant \,. \tag{10}$$

The constant is proportional to the amount of gas.

When the amount of gas is represented by mass (m):
$$p.V = m.R_s.T \,. \tag{11}$$

When the amount of gas is represented by particle number (N):
$$p.V = N.k.T \,. \tag{12}$$

When the amount of gas is represented by the number density (n):
$$p = n.k.T \,. \tag{13}$$

When the amount of gas is represented by the amount of substance (ν):
$$p.V = \nu.R.T \,. \tag{14}$$

The values of the constants in the above equations are as follow:

Boltzmann's constant: $k = 1.38 \times 10^{-23} \, J.K^{-1}$

Universal gas constant: $R = 8.31 \, J.mol^{-1}K^{-1}$

Specific gas constant (depends on the gas species): $R_s = \frac{R}{M}, [R_s] = J.kg^{-1}.K^{-1}$.

This ideal gas law gives the definition of molar volume ($V_m$), which describes the volume of one mole of gas under standard temperature and pressure (STP) conditions (1 atm. and 273 K):

$$V_m = \frac{V}{\nu} = \frac{R.T}{P} = 22.41 m^3.mol^{-1} \,. \tag{15}$$

*3.1.6  Dalton's law*

This states that the total pressure of a gas mixture is the sum of the partial pressures of the individual gases:

$$P_{total} = P_1 + P_2 + \cdots + P_i \,. \tag{16}$$

Considering this, then the ideal gas law can be presented as:
$$P_{total} = (n_1 + n_2 + \cdots + n_i).k.T \,. \tag{17}$$

## 3.2  Kinetic gas theory

*3.2.1  Maxwell−Boltzmann velocity distribution*

When gas molecules collide with each other or with the walls of the container, their velocity and direction change. The gas molecules have a wide range of velocities, however, not all of the velocities have equal probability. The distribution of the particle velocities depends on the mass of the particles and the temperature. The velocity distribution is presented by the Maxwell−Boltzmann velocity distribution [6]:

$$\frac{dn}{dv} = \frac{2N}{\pi^{1/2}} \cdot \left(\frac{m}{2kT}\right)^{3/2} \cdot v^2 \cdot e^{-\left(\frac{mv^2}{2kT}\right)} \,. \tag{18}$$

Where:

*v*: velocity of molecules [m.s$^{-1}$]

*n*: number of molecules with velocity between v and (v + dv)

*N*: the total number of molecules

*m*: mass of molecules [kg]

*K*: Boltzmann constant, 1.38 × 10$^{-23}$ m$^2$ kg s$^{-2}$ K$^{-1}$

*T*: temperature [K]

A presentation of the Maxwell−Boltzmann velocity distribution is shown in Fig. 2.

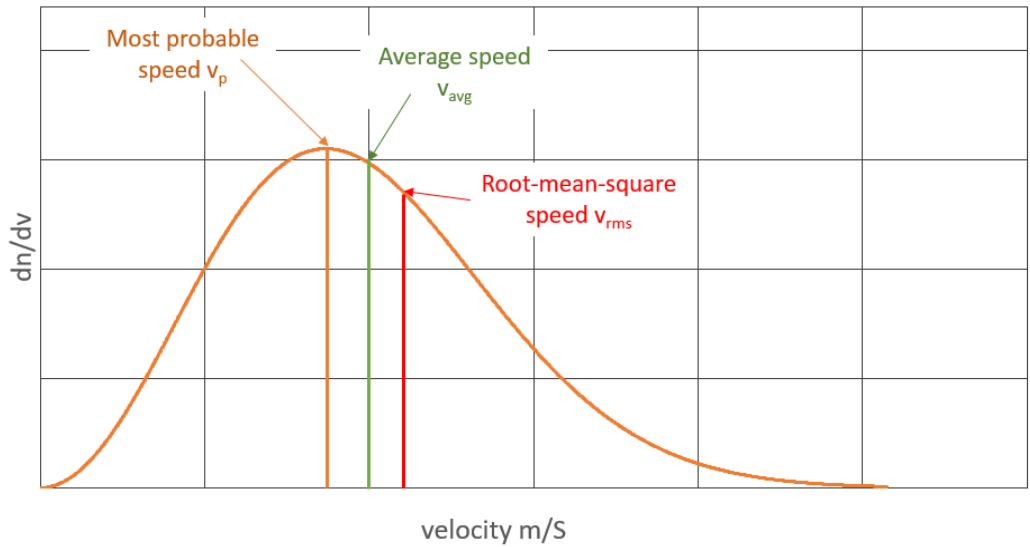

**Fig. 2:** The Maxwell−Boltzmann velocity distribution

As the figure shows, the graph is not symmetrical, it starts with a very small number of molecules with velocities close to zero, and ends with a long tail towards high velocities. For this reason, the average velocity $v_{\text{avg}}$ is slightly shifted towards higher velocities.

The average velocity ($v_{\text{avg}}$) can be described as follow:

$$v_{\text{avg}} = \sqrt{\frac{8kT}{\pi m}}. \tag{19}$$

The most probable velocity ($v_{\text{p}}$), which represents the speed at the peak of the distribution, is:

$$v_{\text{p}} = \sqrt{\frac{2kT}{m}}. \tag{20}$$

And the root mean square velocity ($v_{\text{rms}}$) can be expressed as follow:

$$v_{\text{rms}} = \sqrt{\frac{3kT}{m}}. \tag{21}$$

The dependence on the temperature means that, if the gas has higher temperature, then probably there will be more gas particles with higher velocities. The dependence on the mass of the particles means that, if the gas has high mass, then probably there will be more gas particles with lower velocities.

Fig. 3 shows the Maxwell−Boltzmann velocity distribution for a gas at room temperature and at 500 K and 700 K. As illustrated, when the gas is heated up, the velocities are shifted towards higher velocities and the peak has been reduced in order to have the same area under the curve (same number of gas molecules N).

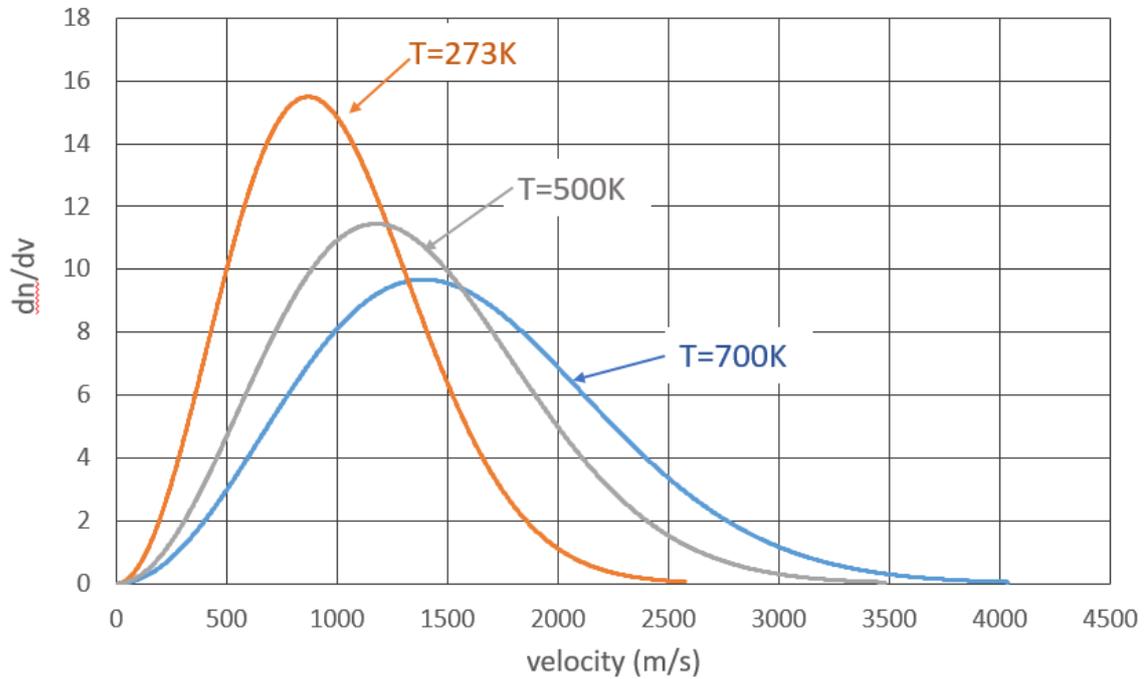

**Fig. 3:** The Maxwell−Boltzmann velocity distribution for a gas system at 273 K, 500 K, and 700 K

Fig. 4 shows the same gas system where the gas has mass (*m*) at room temperature, curve (1), and when the mass is three times higher, curve (2), and an order of magnitude higher, curve (3). The graph shows that heavier gases have lower velocities, while lighter gases have a wide velocity distribution and they have a higher probability of having higher velocities. This explains why helium gas (low mass) is a good candidate to use for leak detection, as it will travel fast from the location of the leak towards the leak detector.

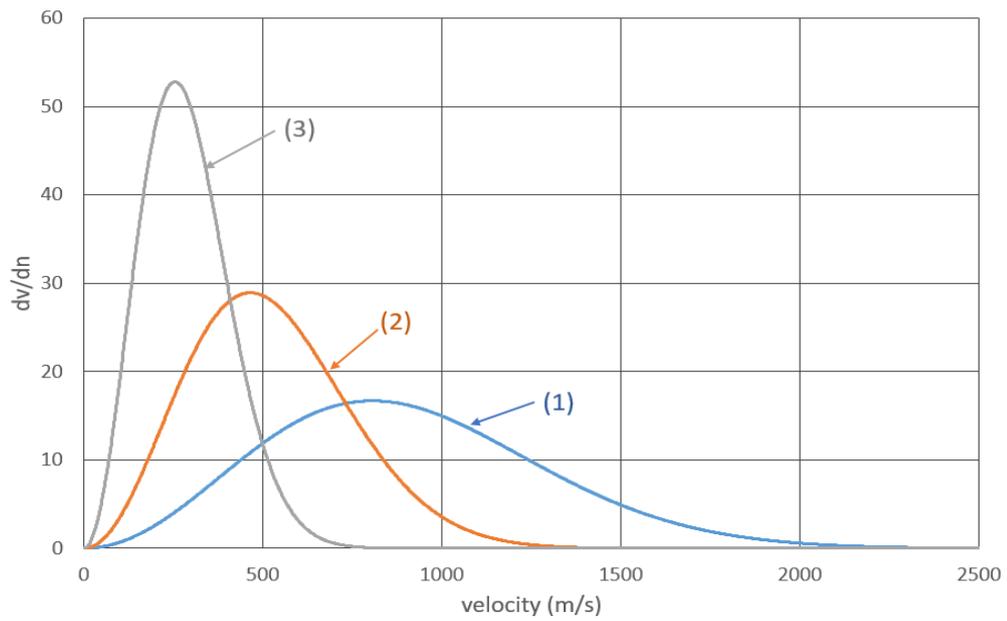

**Fig. 4:** The Maxwell−Boltzmann velocity distribution for a gas system with different masses

### *3.2.2 Mean free path*

The mean free path is defined as the average distance travelled by gas molecules between collisions with the walls of the container or the other particles in the system, see Fig. 5.

The mean free path depends on the gas molecules diameter, the temperature, and the pressure, and is given as follow:

$$\lambda = \frac{k.T}{\sqrt{2}.\pi.d^2.p}. \tag{22}$$

Where:

$\lambda$: mean free path [m]

$d$: molecular diameter [m]

$p$: pressure [Pa]

$K$: Boltzmann constant, $1.38 \times 10^{-23}$ m² .kg.s⁻² .K⁻¹

$T$: temperature [K]

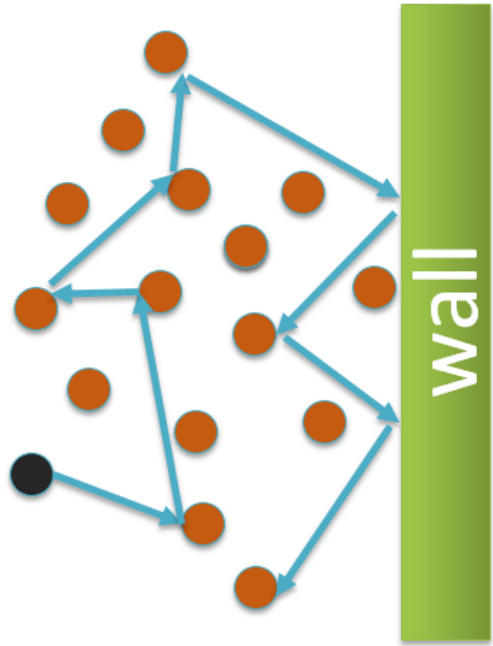

**Fig. 5:** Several collisions occur in the path of a particle, both with the walls and with the other particles

The mean free path in a vacuum is very large; the molecules have a higher probability of colliding with walls of the vacuum chamber rather than with each other. Table 3 shows the mean free path for air at room temperature and various pressure.

**Table 3:** The mean free path of air at room temperature at various pressures

| p [mbar] | $1.0 \times 10^3$ | 1.0 | $1.0 \times 10^{-3}$ | $1.0 \times 10^{-6}$ | $1.0 \times 10^{-9}$ |
|---|---|---|---|---|---|
| λ [cm] | $7.0 \times 10^{-6}$ | $7.0 \times 10^{-3}$ | 7.0 | $7.0 \times 10^3$ | $7.0 \times 10^6$ |

### *3.2.3  Impingement rate, monolayer time, and collision rate*

The impingement rate is defined as the number of particles incident upon a unit surface area per unit time, and is given as:

$$Z_A = \frac{2.635 \cdot 10^{22} \cdot p}{\sqrt{T \cdot M}}. \tag{23}$$

Where:

$Z_A$: impingement rate [cm$^{-2}$. s$^{-1}$]

$M$: molar mass [g.mol$^{-1}$]

$p$: pressure [mbar]

$T$: temperature [K].

From this concept, one can define the time for monolayer formation, which is the time required for a freshly formed surface to become covered with a monolayer of gas molecules.

$$\tau = \frac{n_{mono}}{Z_A}. \tag{24}$$

Where:

τ: monolayer formation time [s]

$Z_A$: impingement rate [cm$^{-2}$. s$^{-1}$]

$n_{mono}$: number of molecules per unit area [cm$^{-2}$].

The typical number of molecules per surface area ($n_{mono}$) is around $1 \times 10^{15}$ molecules.cm$^{-2}$. For air at room temperature, the monolayer time can be expressed as:

$$\tau = \frac{3.2 \cdot 10^{-6}}{p}. \qquad (25)$$

Where $p$: the pressure [mbar].

One can also consider the volumetric collision rate, which is the number of collisions occurring per unit volume per unit time ($Z_V$):

$$Z_V = \frac{5.27 \cdot 10^{22} \cdot p}{\lambda \cdot \sqrt{T.M}}. \qquad (26)$$

Where:

$Z_V$: volume collision rate [cm$^{-3}$. s$^{-1}$]

$M$: molar mass [g.mol$^{-1}$]

$p$: pressure [mbar]

$T$: temperature [K]

$\lambda$: mean free path [cm].

A summary of general gas properties at varies vacuum levels is shown in Table 4 [7]. The gas properties in the UHV range are unique, with pressure in the 10$^{-9}$ mbar range there are around 10 million particles per cm$^3$, the particles will travel for around 100 km before colliding with other particles or with the walls of the vacuum chamber, and a large number of molecules ($1.0 \times 10^{11}$) will impinge on one cm$^2$ of the vacuum walls in one second. Also, around $10^5$ molecules will collide with each other within 1 cm$^3$ in one second, and it will take around three hours at such a low pressure to cover a clean surface with molecules.

Table 4: Approximate gas properties at varies vacuum levels

| vacuum range | pressure (mbar) | gas density (cm$^{-3}$) | mean free path (cm) | impingement rate (cm$^{-2}$.s$^{-1}$) | collision rate (cm$^{-3}$.s$^{-1}$) | monolayer formation time (s) |
|---|---|---|---|---|---|---|
| | p | n | λ | $Z_A$ | $Z_V$ | τ |
| atm. | 10$^3$ | 10$^{19}$ | 10$^{-5}$ | 10$^{23}$ | 10$^{29}$ | $1 \times 10^{-9}$ |
| medium | 1 | 10$^{16}$ | 10$^{-2}$ | 10$^{20}$ | 10$^{23}$ | $1 \times 10^{-6}$ |
| high | 10$^{-3}$ | 10$^{13}$ | 10 | 10$^{17}$ | 10$^{17}$ | $1 \times 10^{-3}$ |
| very high | 10$^{-6}$ | 10$^{10}$ | 10$^4$ | 10$^{14}$ | 10$^{11}$ | 10 |
| ultra-high | 10$^{-9}$ | 10$^7$ | 10$^7$ | 10$^{11}$ | 10$^5$ | 180 |

## 3.3 Gas flow

### 3.3.1 *Gas flow regime*

The production of vacuum or the removal of the gas from inside a vacuum vessel is related to the gas flow. Several gas flow regimes occur when the pressure changes from atmospheric pressure to a very low pressure. The gas flow is governed by the viscosity of the gas when going from atmospheric pressure to a low vacuum, and with further reduction of the pressure, the gas goes into the molecular flow regime, where the flow is governed by the molecular behaviour of the gas.

Fig. 6 shows the different flow regimes. Viscous flow can be turbulent flow or laminar flow. For turbulent flow, the velocity of the gas is high and the gas layers are not flowing in parallel. When the velocity of the gas is lower, then the layers of the gas flow in parallel to each other, with some friction close to the walls of the containers. This flow is called laminar flow. When the gas is in molecular flow, then the molecules are moving randomly. The intermediate regime occurs when the flow follows the molecular behaviour as well as the viscous behaviour.

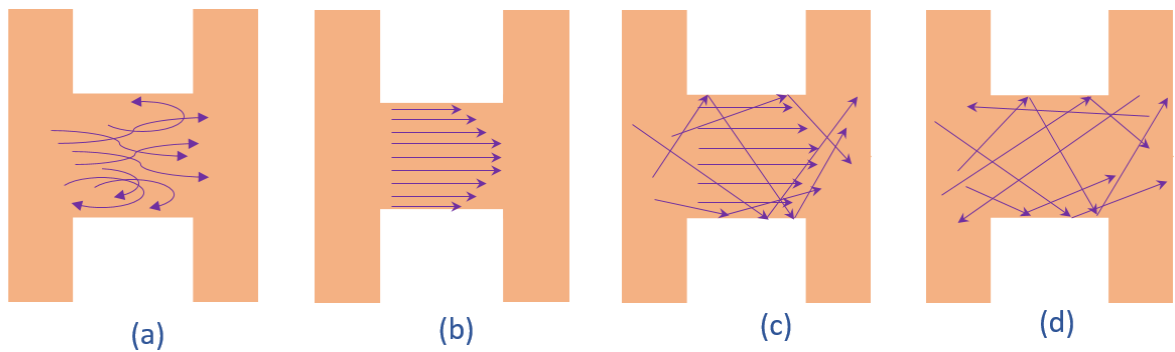

**Fig. 6:** Illustration of the various gas flow regimes through a pipe in a vacuum chamber: viscous flow, which is of two types: (a) turbulent, (b) laminar, intermediate flow (c), and molecular flow (d).

The dimensionless Knudsen number ($K_n$) is used to differentiate between the different flow regimes, and is defined as the ratio of the mean free path to the dimension of the chamber (e.g., dimeter of the pipe (*D*)):

$$K_\mathrm{n} = \frac{\lambda}{D}. \tag{27}$$

- The three flow regimes:
    - Viscous: $K_n < 0.01$.
    - Intermediate: $0.01 < K_n < 0.5$.
    - Molecular: $K_n > 0.5$.

According to the Knudsen number, if the mean free path is larger than the dimension of the chamber (e.g., tube diameter), the gas flow is considered to be molecular flow, where collisions between the gas molecules are less likely to occur and most probably the particles will collide with the walls of the container.

### 3.3.2 *Flow rate*

The flow rate of gas molecules is defined as the transported gas per unit time, and this can be described in different ways [1]:

Volumetric flow rate, $q_V$ [m³.s⁻¹]: $q_\mathrm{V} = \frac{\Delta V}{\Delta t} = \dot{V}$. (28)

Mass flow rate, $q_m$ [kg.s$^{-1}$]: $q_m = \frac{\Delta m}{\Delta t} = \dot{m}$. (29)

Molar flow rate, $q_v$ [mol.s$^{-1}$]: $q_v = \frac{\Delta v}{\Delta t} = \dot{v}$. (30)

Particle flow rate, $q_N$ [s$^{-1}$]: $q_N = \frac{\Delta N}{\Delta t} = \dot{N}$. (31)

### 3.3.3 *Pumping speed and throughput*

When a pump is used to remove the gas from a system, the rate at which the gas is removed is the pumping speed $S$ [l.s$^{-1}$], which is defined as the volume of gas per unit time (volumetric flow rate) which the pump removes from the system at the pressure existing at the inlet of the pump.

$$S = \dot{V}_{inlet} = q_{V,inlet}. \quad (32)$$

Throughput $Q$ or $pV$ flow [mbar.l.s$^{-1}$] is the quantity of gas expressed in pressure–volume ($p.V$) units at a specific temperature flowing per unit time across a specific cross-section:

$$Q = q_{pV} = p.\dot{V}. \quad (33)$$

The relationship between the throughput and the pumping speed at the entrance of the pump can be expressed as:

$$Q = p.S. \quad (34)$$

Fig. 7 shows an example of how the pressure, throughput, and the pumping speed relate to each other in a vacuum system.

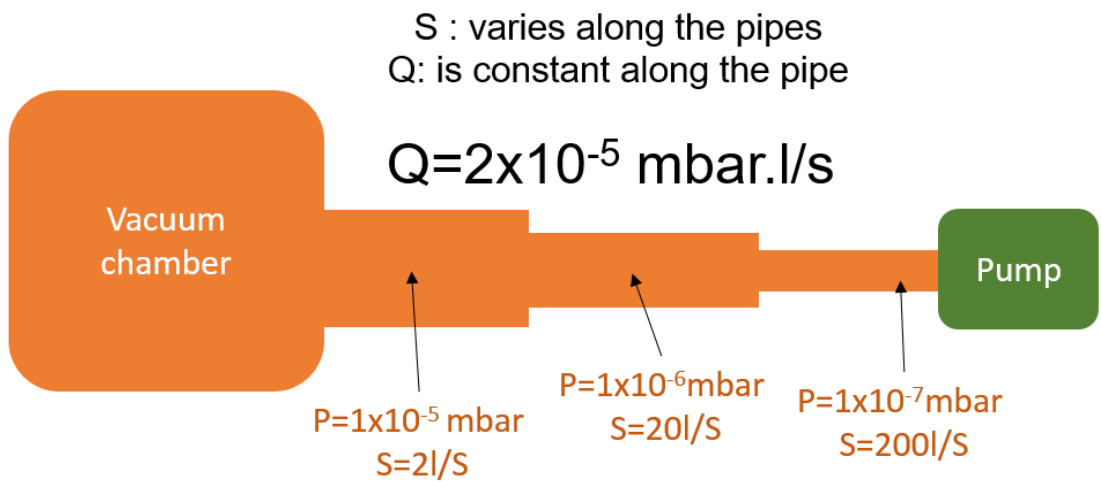

**Fig. 7:** An example of the relation between the pumping speed, throughput, and the pressure of a vacuum system

### 3.3.4 *Conductance*

When the gas flows through a pipe, the pipe will exert a resistance on that flow which is known as the impedance ($Z$). This resistance will cause a pressure drop along the pipe ($p_1$–$p_2$), the reciprocal of the impedance is the conductance $C$ [l.s$^{-1}$]. The gas throughput is directly proportional to the pressure difference, and the conductance $C$ is the constant of proportionality:

$$Q = C(p_1 - p_2). \quad (35)$$

The conductance depends on the temperature, gas species, geometry, and the pressure (in the viscous regime).

A vacuum system is usually made up of several pipes and apertures which are connected to each other in series or in parallel. In order to estimate the pressure profile along the system, one needs to estimate the total conductance of the system.

Fig. 8 shows a vacuum system consisting of a vacuum chamber which is connected to the pump through two pipes connected in series and of different conductance. The estimation of the total conductance of this system is as follows:

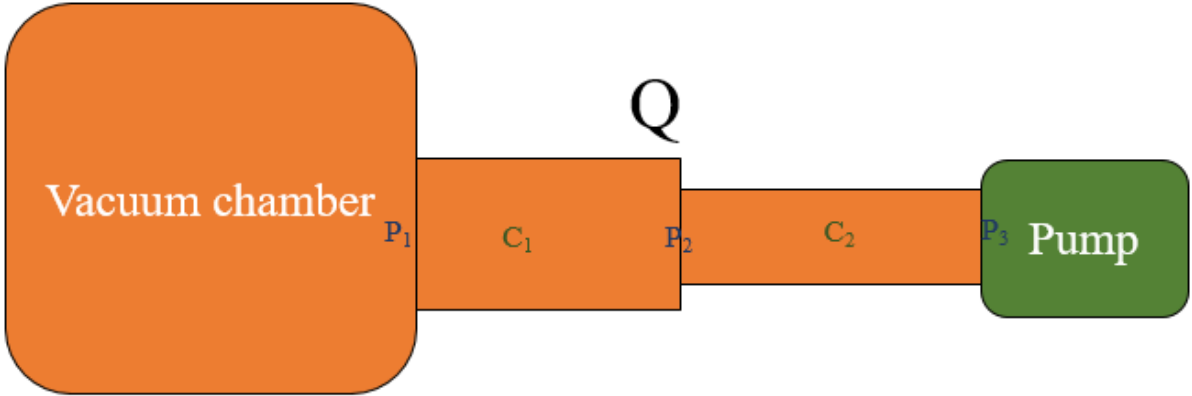

**Fig. 8:** A vacuum system with two tubes connected in series

The throughput through pipe 1:

$$Q_1 = C_1 \cdot (p_1 - p_2). \tag{36}$$

The throughput through pipe 2:

$$Q_2 = C_2 \cdot (p_2 - p_3). \tag{37}$$

However:

$$Q_1 = Q_2 = Q = C_{\text{total}}(p_1 - p_3). \tag{38}$$

Then:

$$p_1 - p_2 = \frac{Q}{C_1} \text{ and } p_2 - p_3 = \frac{Q}{C_2}. \tag{39}$$

$$p_1 - p_3 = \frac{Q}{C_{\text{total}}} = \frac{Q}{C_1} + p_2 + \frac{Q}{C_2} - p_2 = Q\left(\frac{1}{C_1} + \frac{1}{C_2}\right). \tag{40}$$

Accordingly, the total conductance for a system of pipes connected in series is:

$$\frac{1}{C_{\text{total}}} = \frac{1}{C_1} + \frac{1}{C_2} = \sum_i^N \frac{1}{C_i}. \tag{41}$$

Fig. 9 shows a vacuum system that consists of a vacuum chamber which is connected to the pump through two pipes in parallel and of different conductance. The estimation of the total conductance of this system is as follows:

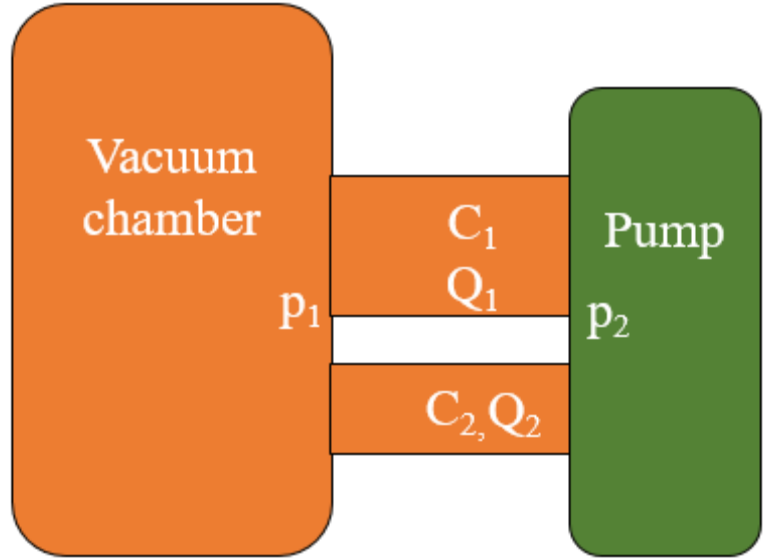

**Fig. 9:** A vacuum system made of two pipes connected in parallel

The throughput through pipe 1:
$$Q_1 = C_1 \cdot (p_1 - p_2). \tag{42}$$

The throughput through pipe 2:
$$Q_2 = C_2 \cdot (p_1 - p_2). \tag{43}$$

The total throughput of the system is:
$$Q_{\text{total}} = Q_1 + Q_2 = C_{\text{total}}(p_1 - p_2). \tag{44}$$

Then:
$$Q_{\text{total}} = C_{\text{total}}(p_1 - p_2) = C_1 \cdot (p_1 - p_2) + C_2 \cdot (p_1 - p_2). \tag{45}$$

Accordingly:
$$C_{\text{total}} = C_1 + C_2 = \sum_i^N C_i. \tag{46}$$

### 3.3.4.1 *Examples of conductance in the molecular regime*

If two vacuum chambers are connected to each other through a small thick orifice of cross-section $A$, then the conductance of that orifice is expressed as follow:

$$C = 3.64 \cdot A \cdot \sqrt{\frac{T}{M}}. \tag{47}$$

For air at 23°C, the conductance of an orifice is:

$$C = 11.6 \cdot A. \tag{48}$$

For a long circular tube of dimeter $d$ and length $L$, the conductance is as follow:

$$C = 3.81 \cdot \frac{d^3}{L} \cdot \sqrt{\frac{T}{M}}. \qquad (49)$$

For air a 23°C, the conductance of a long circular tubes is:

$$C = 12.1 \times \frac{d^3}{L}. \qquad (50)$$

Where:

$C$: the conductance [l.s$^{-1}$]

$M$: molar mass [g.mol$^{-1}$]

$A$: area [cm$^2$]

$T$: temperature [K]

L: tube length [cm]

$d$: tube diameter [cm].

### 3.3.5　*The effective pumping speed of a pump*

Fig. 10 shows a vacuum chamber which is connected to a pump that has a nominal pumping speed ($S_{nom}$) through a long tube of conductance $C$. The effective pumping speed ($S_{eff}$) at the exit of this pipe to the system is reduced due to the conductance of the pipe; accordingly, the effective pumping speed can be expressed as follows:

$$\frac{1}{S_{eff}} = \frac{1}{C} + \frac{1}{S_{nom}}. \qquad (51)$$

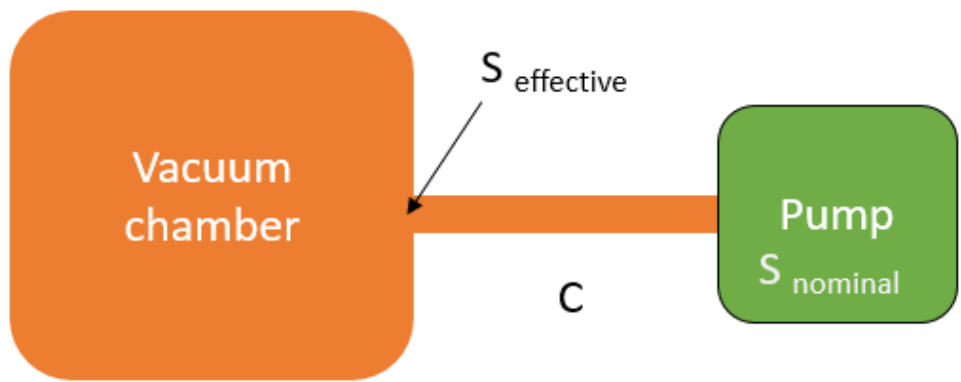

**Fig. 10:** The effect of the conductance of a tube on the effective pumping speed of a pump

# 4 Introduction to the gas sources in particle accelerators

The gas can find its way to the vacuum system through several processes, whose definitions are as follows [8]:

(i) Outgassing: is the spontaneous release of the gas from a solid or liquid.

(ii) Degassing: deliberate removal of the gas from a solid or liquid.

(iii) Desorption: release of the adsorbed species from the surface of a solid or liquid.

## 4.1 Gas sources in vacuum systems

In addition to the residual gas which may stay inside the vacuum chamber after pumping down, there are other gas sources inside a vacuum system. Fig. 11 shows the possible sources of gas inside a vacuum system:

(i) Back-streaming: the gas could go back from the pump to the vacuum system.

(ii) Real leaks: the chamber may have cracks, bad welds or sealing, etc., all of which may allow the gas from the outside to pass through those leaks into the vacuum system.

(iii) Virtual leaks: a trapped volume of gas which is connected to the vacuum chamber via a small path with small conductance. The trapped volume could be caused by un-vented screws inside the vacuum, incorrect welding, etc.

(iv) Desorption: gas molecules are bound to the inner surface of the chamber (e.g., when venting has occurred), and if they have enough energy they will leave or desorb into the vacuum surface.

(v) Permeation: the gas can permeate from outside to the bulk of the material. The gas then dissolves in the bulk and diffuses into the inner walls of the chamber and then desorbs into the vacuum system. An example of this is elastomer seals, which have a permeation level which is high compared to metal.

(vi) Bulk diffusion: the gas which is trapped inside the chamber bulk could travel between the metal grain boundaries into the vacuum chamber.

(vii) Evaporation: if the material inside the vacuum system has a high vapour pressure, then once the system is evacuated and the pressure is lower than the material vapour pressure, that material will evaporate and find its way into the vacuum system. For this reason, materials such as lead and zinc, which have a high vapour pressure, should be avoided for use in UHV.

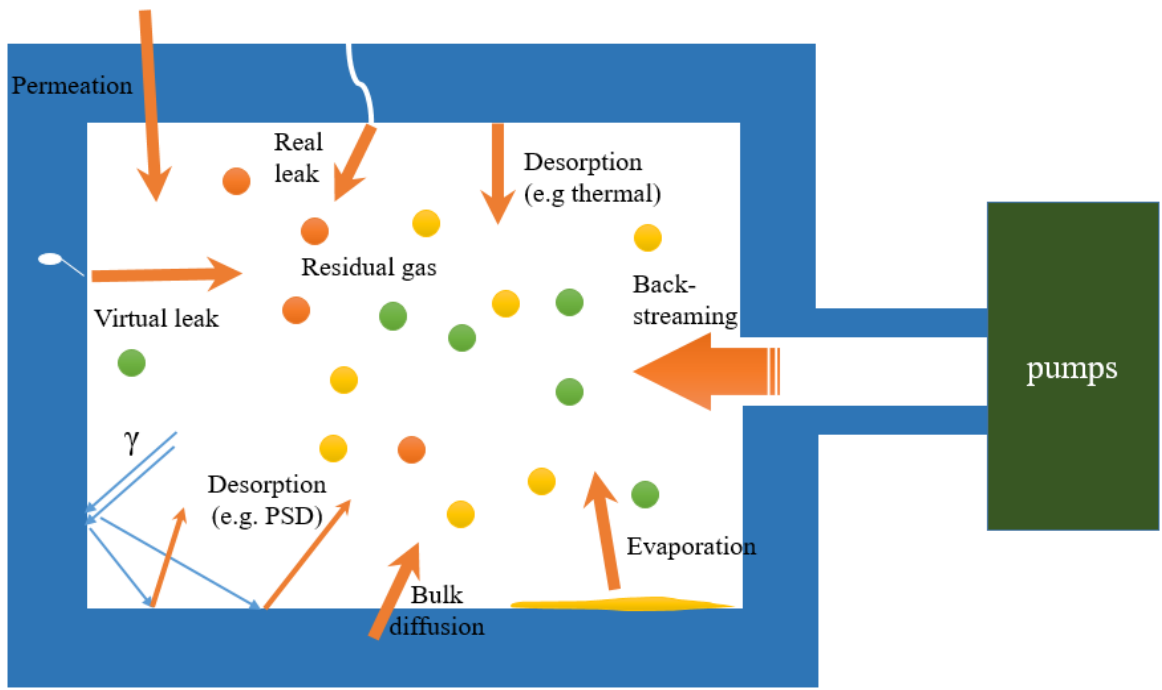

**Fig. 11:** Possible sources of gas inside a vacuum chamber

## 4.2 Main sources of gases in particle accelerators

In particle accelerators, the main sources of gas are due to:

(i) Thermal outgassing: which occurs due to bulk diffusion as well as due to desorption from the surface, for example when the chamber is pumped down, see Fig. 12. The thermal outgassing depends on the material, cleaning, history, pump down time, etc.

(ii) Beam induced desorption (BID): this occurs when the outgassing is stimulated by the impact of energetic particles, such as photons, ions, or electrons, which are created from the operation of the particle accelerators.

The beam induced desorption is characterized by the desorption yield ($\eta$):
$$\eta = \frac{\text{number of desorbed molecules}}{\text{number of particles impinging the surface}}. \tag{52}$$

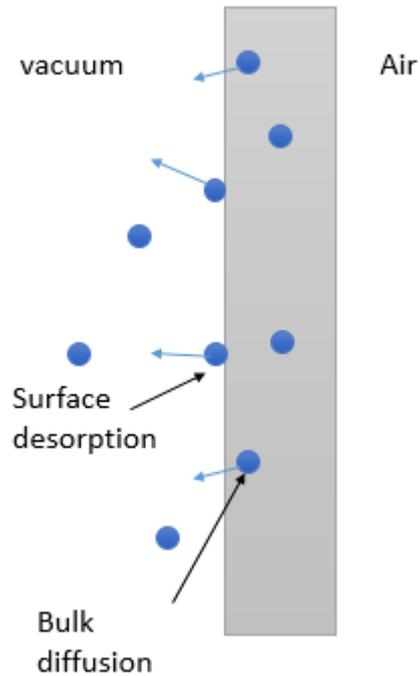

**Fig. 12:** The thermal outgassing process

The BID depends on many parameters:

(i) Incident particle: type and energy.

(ii) Material.

(iii) Surface roughness.

(iv) Cleanliness of the surface.

(v) History of the material.

(vi) Temperature.

(vii) Integrated particle dose.

Examples of BID are: photon-stimulated desorption, ion-induced desorption, and electron-induced desorption.

## 4.2.1 Photon-stimulated desorption

Photon-stimulated desorption (PSD) occurs when the energized photons impinge on the vacuum chamber walls, which causes electron emission, called photo-electrons. When the photo-electrons leave the surface, they will desorb molecules, and also when they hit the surface, they will also desorb gas molecules, see Fig. 13.

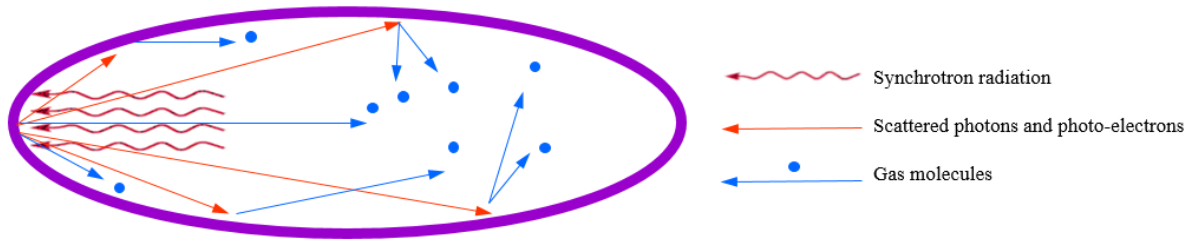

**Fig. 13:** The process of PSD

PSD yield is defined as the number of molecules desorbs for each photon impinging the surface.

Experiments were performed to measure the PSD yield and its evolution with the beam dose for various materials, incident angles, and beam energies. An example of the evolution of the PSD yield with the beam dose is shown in Fig. 14, where the chamber is made of stainless steel baked to 200°C for 48 h., and beam energy is 500 eV [9].

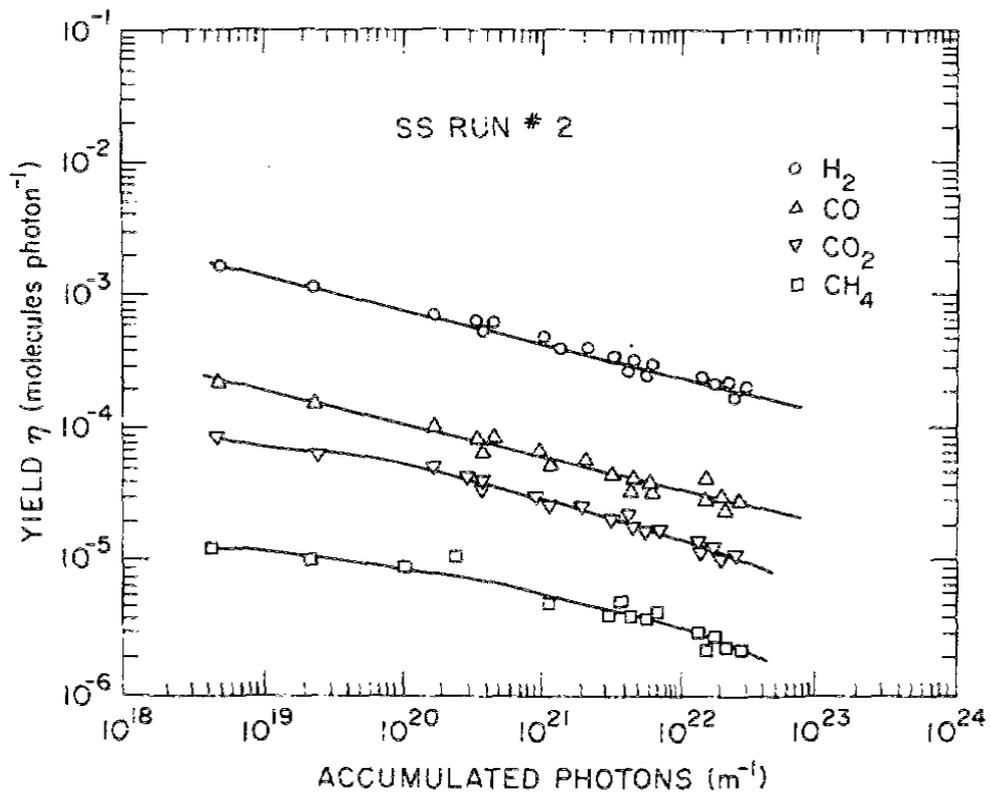

**Fig. 14:** The PSD yield of in situ baked stainless steel vacuum chamber

*4.2.2  Ion-induced desorption*

In positively charged rings, when the circulating beam hits the residual gas inside the vacuum chamber, the beam will ionize the residual gas, the ionized residual gas will be repelled from the positively charged beam, and subsequently, the ions will hit the walls of the vacuum chamber. Ions can gain energies which are effective in desorbing the molecules bound to the walls of the vacuum chamber, and now there is more residual gas and hence more ionization will occur, resulting in a continuous increase in the pressure. This cycle will continue, generating a high outgassing and a high pressure inside the system. Figure 15 shows an illustration of the process of ion-induced desorption.

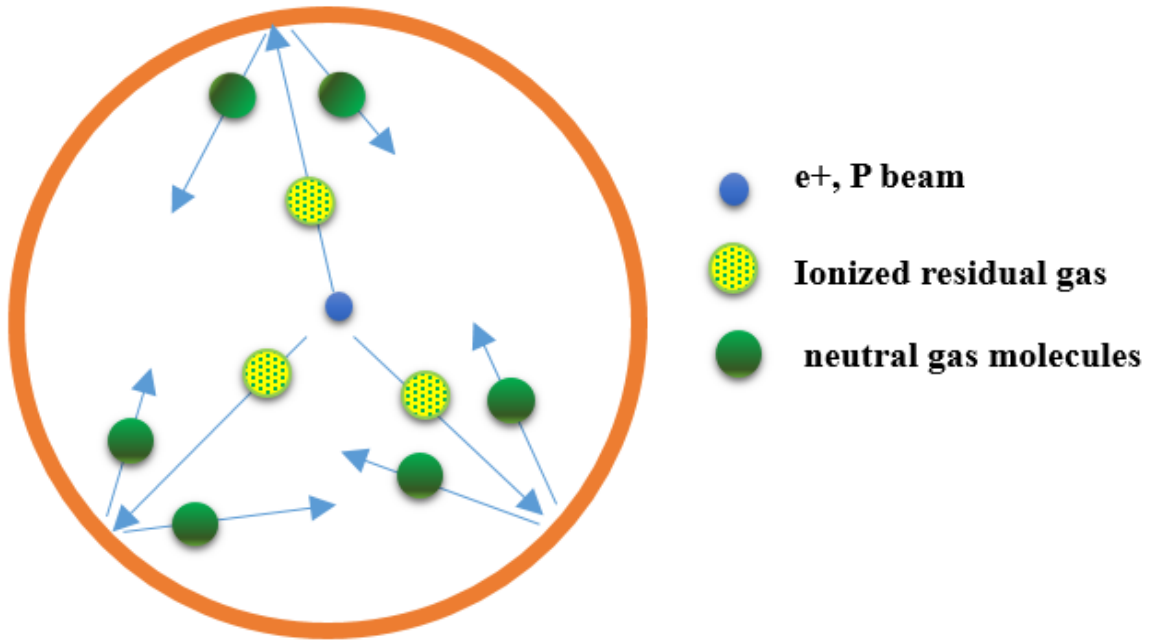

**Fig. 15:** Illustration of the ion-induced desorption process

*4.2.3  Electron-induced desorption (EID)*

When high energy electrons bombard the surface of the chamber, they can excite the electrons of the molecules which are weakly bound to that surface and this may break bonds, allowing the atoms to desorb as ions or neutrals.

EID yield is defined as the number of molecules desorb from the chamber wall for each electron impinging the surface.

The electron-induced desorption EID yield and photon-stimulated desorption yield are coupled, as photo-electrons are part of the photon-stimulated gas desorption process. Experiments were performed to measure the EID yield and its evolution and dependence on the beam dose, material, temperature, cleaning effect, beam energy, etc. An example of the evolution of the EID yield with the beam dose is shown in Fig. 16 for a sample of stainless steel 316 LN at 36°C which has been pre-baked for 24 h at 150°C and degassed for 2 h at 300°C. The electron beam energy was 300 eV [10].

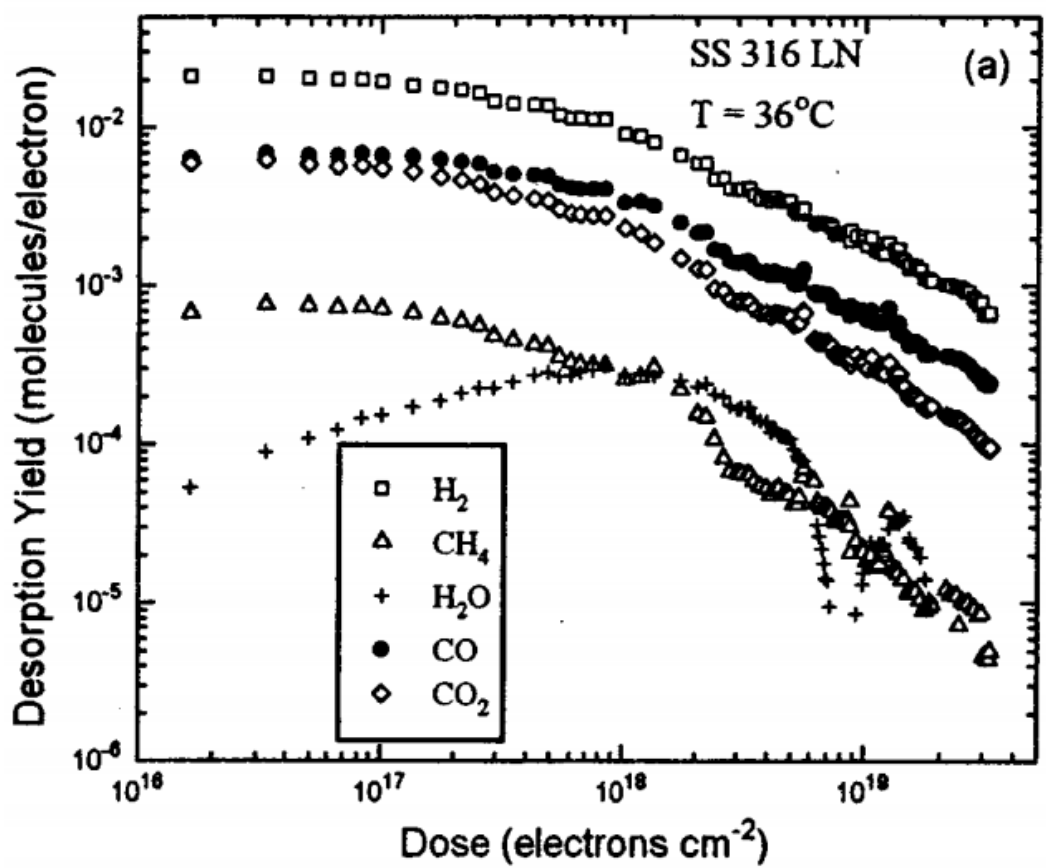

**Fig. 16:** The electron induced desorption yield for stainless steel 316 LN as a function of the accumulated beam dose at 36°C.